\documentclass[epj,twocolumn]{webofc} 
\usepackage[varg]{txfonts}   
\usepackage{graphicx,amsmath,wrapfig,epsfig}
\woctitle{ISVHECRI 2018}

\usepackage{color}

\definecolor{mycolor}{rgb}{0.6,0.0,0.4}

\begin{document}
\title{High-energy neutrino-nucleus interactions}
\author{S. Kumano\inst{1,2}} 
\institute{
KEK Theory Center, Institute of Particle and Nuclear Studies,
KEK, \\
\ \, 1-1, Ooho, Tsukuba, Ibaraki, 305-0801, Japan
\and
J-PARC Branch, KEK Theory Center, Institute of
Particle and Nuclear Studies, KEK, \\
\ \, and Theory Group, Particle and Nuclear Physics Division, 
  J-PARC Center, \\
\ \, 203-1, Shirakata, Tokai, Ibaraki, 319-1106, Japan}

\abstract{
High-energy neutrino-nucleus interactions are discussed by considering
neutrino-oscillation experiments and ultra-high-energy cosmic neutrino
interactions. The largest systematic error for the current neutrino 
oscillation measurements comes from the neutrino-nucleus interaction 
part, and its accurate understanding is essential for high-precision
neutrino physics, namely for studying CP violation in the lepton sector.
Depending on neutrino beam energies, quasi-elastic, resonance, Regge,
or/and deep inelastic scattering (DIS) processes contribute to 
the neutrino cross section. It is desirable to have a code to 
calculate the neutrino-nucleus cross section in any kinematical
range by combining various theoretical descriptions.
On the other hand, the IceCube collaboration started obtaining
cross section data up to the $10^{15}$ eV range, so that 
it became necessary to understand ultra-high-energy neutrino 
interactions beyond the artificial lepton-accelerator energy range.
For future precise neutrino physics including the CP measurement,
it is also necessary to understand accurate nuclear corrections.
The current status is explained for nuclear corrections in
DIS structure functions.
The possibility is also discussed to find gravitational sources
within nucleons and nuclei, namely matrix elements of quark-gluon
energy-momentum tensor. They could be probed by neutrino interactions
without replying on direct ultra-weak ``gravitational interactions'' 
with high-intensity neutrino beams, possibly 
at a future neutrino factory,
by using techniques of hadron tomography.
}

\maketitle

\section{Introduction}
\label{intro}

In recent years, precise descriptions of high-energy neutrino-nucleon 
and neutrino-nucleus reactions became necessary. 
First, it is motivated by the progress of neutrino oscillation 
experiments. The measurements are getting more and more precise, 
and it became the stage of probing CP violation in leptons.
The target is water, for example, in the T2K experiment, so 
that accurate neutrino-oxygen cross sections should be 
calculated in addition to neutrino-nucleon ones
\cite{neutrino-nucleus}.
Roughly, the accuracy of 5\% is required for the theoretical
estimate on the neutrino-nucleus cross sections
for future neutrino-oscillation measurements.

Second, ultra-high-energy neutrino experiments have been
done recently by the IceCube collaboration.  
In fact, the first cross section measurement was reported
in the $10^{13} - 10^{15}$ eV range by the IceCube in 2017
\cite{icecube}.
So far, the IceCube cross sections are consistent with 
the standard model estimate within their errors.
In the near future, we expect to have much experimental progress 
in this field by considering other future plans on high-energy 
neutrino experiments such as
KM3NeT (Cubic Kilometer Neutrino Telescope)
and Baikal GVD (Gigaton Volume Detector).

Considering these circumstances, we explain descriptions
of high-energy neutrino-nucleon and neutrino-nucleus 
cross sections by focusing mainly 
on the deep-inelastic-scattering (DIS) part.
The current status and future prospect of
structure functions and parton distribution functions
are summarized in Ref.\,\cite{dis-kumano}.
In Sec.\,\ref{neutrino-nucleus}, outline of neutrino-interaction
descriptions is discussed, and the details are explained
for neutrino DIS processes in Sec.\ref{dis-nucleus}, 
together with nuclear corrections.
The current status of cosmic ultra-high-energy neutrino
cross sections is explained in Sec.\,\ref{cosmic-neutrino}.
Nuclear modifications of the structure function $F_2$ and the PDFs
are shown in Sec.\,\ref{nuclear-mod}.
It is possible to probe gravitational form factors of hadrons
by neutrino scattering as explained in Sec.\,\ref{gravitational}.
The current situation for determining the gravitational form factors
from the KEKB measurements is also shown. 
These discussions are summarized in Sec.~\ref{summary}.

\vspace{-0.20cm}
\section{Neutrino-nucleus interactions}
\label{neutrino-nucleus}
\vspace{-0.10cm}

\begin{figure*}[t]
\hspace{0.2cm}
\begin{minipage}{0.48\textwidth}
  \begin{center}
     \includegraphics[width=6.0cm]{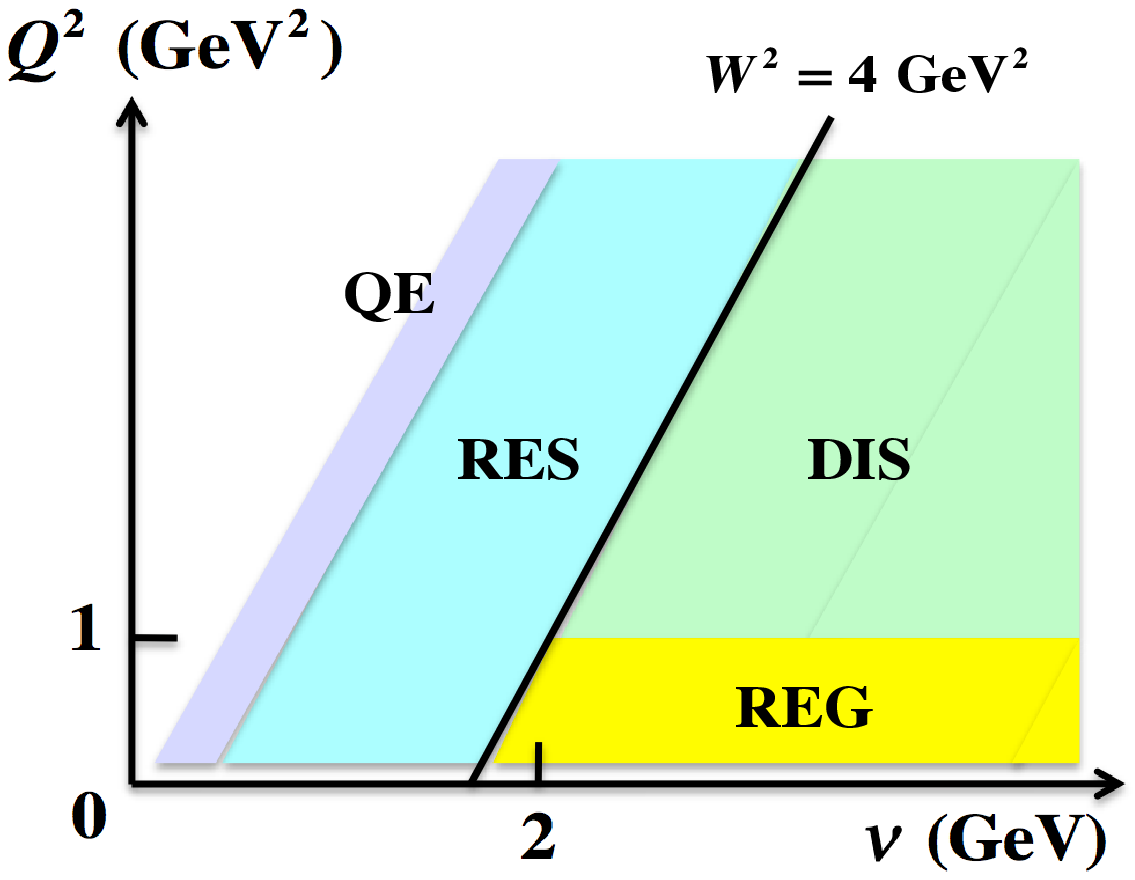}
  \end{center}
\vspace{-0.6cm}
\caption{Kinematical regions of neutrino-nucleus scattering.}
\label{fig:nu-a-kinematics}
\end{minipage}
\hspace{1.2cm}
\begin{minipage}{0.48\textwidth}
   \vspace{0.35cm}
   \hspace{-0.00cm}
     \includegraphics[width=6.0cm]{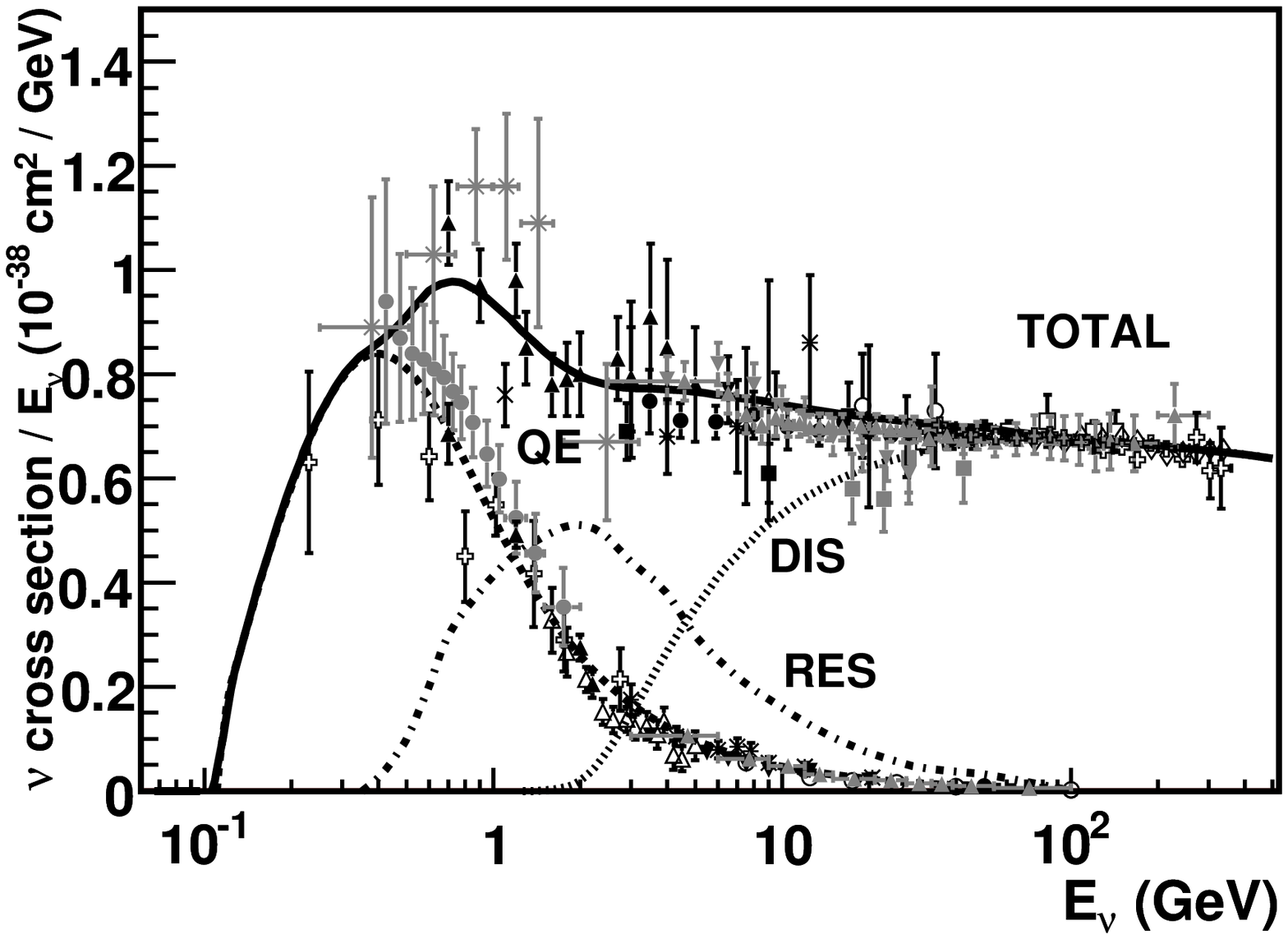}
\vspace{-0.30cm}
\caption{
Neutrino scattering cross sections \cite{nu-each}
.}
\label{fig:nu-cross}
\end{minipage} 
\vspace{-0.7cm}
\end{figure*}

Kinematical regions of neutrino-nucleus scattering are
shown in Fig.\,\ref{fig:nu-a-kinematics} by considering
the neutrino energies from MeV to multi GeV, namely
current beam energies of neutrino oscillation experiments.
There are four kinematical regions as shown in 
Fig.\,\ref{fig:nu-a-kinematics}: quasi-elastic (QE), resonance (RES), 
deep inelastic scattering (DIS), and Regge (REG).
Here, four-momentum and energy transfers are denoted 
$q$ and $\nu$, respectively, and $Q^2$ is defined by
$Q^2 = -q^2$.

The elastic scattering occurs if the relation
$\nu = Q^2 /(2 M_A)$, where $M_A$ is the nuclear mass,
is satisfied. At low energy neutrino scattering, this relation
is given approximately by $\nu \simeq \vec q^{\,2} /(2 M_A)$.
It means that the energy transfer is equal to the recoil energy
of a nucleus. In this elastic scattering, the neutrino sees
the nucleus as a whole system.
As the neutrino energy increases, 
the Compton wavelength becomes short enough to resolve 
individual nucleons in the nucleus. 
If the nucleon is at rest, the kinematical 
relation becomes
$\nu = Q^2 /(2 M_N) \simeq \vec q^{\,2} /(2 M_N)$.
Then, we take into account binding and Fermi-motion effects.
The binding effect could be effectively included into 
a modified nucleon mass $M_N^{\, *}$ as a rough estimate,
so that the kinematical relation is expressed by
the nucleon momentum $\vec p$ as
$\nu \simeq (\vec p + \vec q \, )^{2} /(2 M_N^{\, *})
            - \vec p^{\,2} /(2 M_N^{\, *})$.
By using the the nucleon's Fermi momentum $p_{_F}$,
it is expressed as        
$\vec q^{\,2} - 2 \, q \, p_{_F} \le 2 M_N^{\, *} \nu
 \le \vec q^{\,2} + 2 \, q \, p_{_F}$
 \cite{lepton-kinematics}.
This is the quasi-elastic region shown in Fig.\,\ref{fig:nu-a-kinematics}.

As the neutrino energy increases further, 
the nucleon is excited into resonances,
and this kinematical region is given by 
the invariant-mass squared as
$M_N^2 < W^2 = (p+q)^2 < 4$ GeV$^2$. This is the
resonance region. 
At $W^2 = (p+q)^2 \ge 4$ GeV$^2$ and $Q^2 \ge 1$ GeV$^2$, 
the nucleon is broken up into hadron pieces, and it is 
the deep-inelastic-scattering (DIS) region. 
The neutrino scattering in this region is described
by partons within the nucleus.
Because the photon interacts with the partons within
a short time, interactions among the partons could be neglected
for describing the process. Namely, the reaction is
described by the incoherent impulse approximation 
from individual partons.
For the parton description with the incoherent assumption,
the $Q^2$ value should be large enough. Furthermore, 
$Q^2$ dependencies of structure functions or parton 
distribution functions (PDFs) are described by 
the Dokshitzer-Gribov-Lipatov-Altarelli-Parisi (DGLAP)
evolution equations \cite{dglap-mk}
in the perturbative QCD region. 
The running coupling constant $\alpha_s (Q^2)$ should be
small enough for the perturbative calculations,
and it means $Q^2$ is large enough:
$Q^2 \ge 1$ GeV$^2$ or a few GeV$^2$.
There is another region at $W^2 \ge 4$ GeV$^2$ 
and $Q^2 < 1$ GeV$^2$, and the reaction is described
by the Regge theory with Reggeons and Pomerons.
There are also guidelines from the quark-hadron duality
and 
the partially conserved axial current (PCAC) 
in the limit $Q^2 \to 0$ for describing
the axial-vector current part, whereas the vector part is conserved.

In Fig.\,\ref{fig:nu-cross}, 
each contribution is shown as a function of neutrino energy
\cite{nu-each}. At low energies with $E_\nu <1$ GeV such as T2K, 
the reactions are dominated by the quasi-elastic scattering
and there are some resonance effects.
On the other hand, the DIS processes started to contribute
in the multi-GeV region of Fermilab neutrino energies.
All the different kinematical regions need to be understood 
precisely for neutrino-oscillation measurements,
and the current situation is summarized in Ref.\,\cite{neutrino-nucleus}.
The ultra-high energy cosmic neutrino reactions are dominated
by the DIS in general, except for the forward scattering.
However, the energy range is beyond our current experimental
understanding by lepton accelerators, so that theoretical
extrapolations are needed for calculating neutrino cross sections
at very high energies.

\vspace{-0.10cm}
\section{Deep inelastic neutrino-nucleus scattering}
\label{dis-nucleus}
\vspace{-0.10cm}

\subsection{Neutrino cross sections and structure functions}
\label{dis-cross}
\vspace{-0.10cm}

In this article, we do not step into the low-energy descriptions
of neutrino-nucleus interactions. One may look at the summary
article \cite{neutrino-nucleus} for such information.
Here, we focus on the DIS part.
The charged-current (CC) cross section for neutrino (or antineutrino)
proton scattering is expressed by three structure functions,
$F_1$ (or $F_L$), $F_2$, and $F_3$ as 
\cite{nu-dis-cross}
\begin{align}
\frac{d\sigma_{CC}^{\nu/\bar\nu }}{dx \, dy}
& =  \frac{G_F^2 \, s }{2 \pi \, (1 + Q^2 /M_W^2 )^2} 
     \, \bigg [  \, F_1^{\, CC} \, x \, y^2
\nonumber \\
& \! \! \! \! \! \! \! \! \! 
     + F_2^{\, CC} \, \left ( 1 -y -\frac{Mxy}{2E} \right )
     \pm F_3^{\, CC} \, xy \, 
         \left( 1 - \frac{y}{2} \right ) \,
        \, \bigg ] ,
\end{align}
where 
$x$ is the Bjorken scaling variable defined by $x=Q^2 /(2 p \cdot q)$,
$\pm$ indicates $+$ and $-$ for neutrino and antineutrino,
respectively, $G_F$ is the Fermi coupling constant,
$y$ is defined by $y = p \cdot q / p \cdot k = \nu /E$
with the neutrino momentum $k$ and energy $E$,
$M_W$ is the $W$ boson mass,
and $s$ is the center-of-mass energy.
In parton model, the structure function $F_1$ is related to $F_2$ 
by the Callan-Gross relation 
$2x F_1^{CC} = F_2^{CC}$, and
the structure functions are expressed by the PDFs for the proton as
\begin{align}
F_2^{\nu p \, (CC)} 
    & = 2x \, \big ( \, d + s + \bar u + \bar c \, \big ) ,
\nonumber \\
x F_3^{\nu p \, (CC)} 
    & = 2x \, \big ( \, d + s - \bar u - \bar c \, \big ) ,
\nonumber \\
F_2^{\bar\nu p \, (CC)} 
    & = 2x \, \big ( \, u + c + \bar d + \bar s \, \big ) ,
\nonumber \\
x F_3^{\bar\nu p \, (CC)} 
    & = 2x \, \big ( \, u + c - \bar d - \bar s \, \big ) .
\label{eqn:f123-parton}
\end{align}

On the other hand, the neutral-current (NC) cross sections are expressed 
in the similar way as
\vspace{-0.20cm}
\begin{align}
\frac{d\sigma_{NC}^{\nu/\bar\nu \, p}}{dx \, dy}
& =  \frac{\rho \, G_F^2 \, s }{2 \pi \, (1 + Q^2 /M_Z^2 )^2} 
      \bigg [  \, F_1^{\, NC} \, x \, y^2
\nonumber \\
& \! \! \! \! \! \! \! \! \! 
 + F_2^{\, NC} \, \left ( 1 -y -\frac{Mxy}{2E} \right )
    \pm F_3^{\, NC} \, xy \, 
                    \left( 1 - \frac{y}{2} \right ) \,
              \, \bigg ] ,
\label{eqn:dis-cross-2-nu-nc}
\\[-0.70cm]
\nonumber
\end{align}
where 
$M_Z$ is the Z-boson mass, $\sin \theta_W$ is the weak-mixing angle,
and $\rho = M_W^2 / \big ( M_Z^2 \, \cos^2\theta_W \big )$.
In the parton model, $F_1$ is related to $F_2$ as
$2x F_1^{NC} = F_2^{NC}$, and
the proton structure functions are expressed by the PDFs as
\vspace{-0.20cm}
\begin{align}
\! \! 
F_2^{\nu/\bar\nu\, p \, (NC)} 
  & \! = \! 2x \, \bigg [ \, (u_L^2+u_R^2) \, \big ( u^+ + c^+ \big )
      + (d_L^2+d_R^2) \, \big ( d^+ + s^+ \big ) \, \bigg ] ,
\nonumber \\
\! \!  
x F_3^{\nu/\bar\nu\, p \, (NC)} 
  & \! = \! 2x \, \bigg [ \, (u_L^2-u_R^2) \, \big ( u^- + c^- \big )
      + (d_L^2-d_R^2) \, \big ( d^- + s^- \big ) \, \bigg ] ,
\\[-0.80cm]
\nonumber
\end{align}
where $q^\pm$ is defined by $q^\pm \equiv q \pm \bar q$.
The left- and right-hand couplings for a quark are 
expressed by the third component of isospin $T_a^{\, 3}$,
charge $e_q$, and the weak-mixing angle $\theta_W$ as
\vspace{-0.30cm}
\begin{align}
&
q_{_L} =  T_q^{\, 3} - e_q  \sin ^2  \theta _W , \ \ 
q_{_R} =  - e_q  \sin ^2  \theta _W ,
\nonumber \\
& \ \ \ \ \ 
T_q^{\, 3},\ e_q  =
\begin{cases}
    +1/2, \ +2/3    & \text{for } q=(u,c) \\
    -1/2, \ -1/3    & \text{for } q=(d,s)
\end{cases} .
\\[-0.70cm]
\nonumber
\end{align}
The parton-model expressions are given in the leading order of 
the running coupling constant $\alpha_s$. Higher-order corrections
are included in coefficient functions together with the gluon
distribution.

\vspace{-0.10cm}
\subsection{Ultra-high-energy cosmic neutrino interactions}
\label{cosmic-neutrino}
\vspace{-0.10cm}

The structure functions have been measured mainly in charged-lepton
DIS processes for nucleons and nuclei. There are also experiments
in neutrino DIS particularly by the NuTeV collaboration.
The highest-energy measurements by artificial accelerators 
were done at HERA (Hadron-Electron Ring Accelerator)
for the nucleon structure functions.
However, there are ultra-high-energy cosmic neutrinos to reach to earth,
and they are detected by the IceCube facility at the antarctic.
The ultra-high-energy neutrino reactions provide valuable information
on nucleon structure studies in the energy region beyond the current
artificial accelerators, 
new physics beyond the standard model,
astrophysical neutrino production, and 
propagation in the cosmic microwave background,
and Greisen-Zatsepin-Kuzmin (GZK) mechanism.

\begin{figure}[t]
    \hspace{+0.30cm}
 \begin{minipage}[c]{0.45\textwidth}
    \vspace{+0.20cm}
    \hspace{+0.80cm}
    \includegraphics[width=5.0cm]{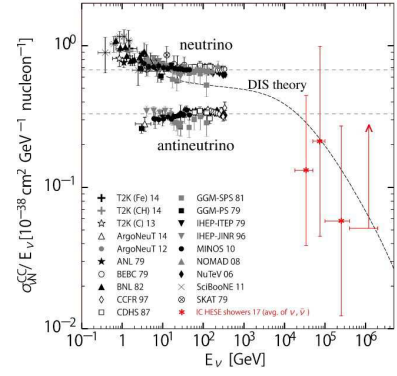}
\vspace{-0.45cm}
\caption{Ultra-high-energy neutrino cross sections \cite{icecube}
(Figure supplied by M. Bustamante).}
\label{fig:icecube-cross}
\vspace{-0.7cm}
 \end{minipage}
\end{figure}

The ultra-high-energy comic neutrino measurements have been
done by the IceCube collaboration. Recently, there was 
a first report on neutrino cross sections \cite{icecube}.
Cosmic neutrinos pass through the earth interior and interact
with the ice surrounding the IceCube detectors in the ice cap
at 1,450-2,450 meters below its surface.
Their results are shown in Fig.\,\ref{fig:icecube-cross}
for the charged-current (CC) cross sections
in the energy range of $10^{13}$-$10^{15}$ eV
together with artificial accelerator measurements
at lower energies, separately, for neutrinos and antineutrinos.
The neutrino CC interactions in the ice 
create charged muons, which are detected as long tracks of
Cherenkov light. However, they are not able to distinguish
neutrino events from antineutrino ones and they are also not
able to detect neutral current events.
In this way, the CC cross sections were
obtained in the IceCube experiment as shown 
in Fig.\,\ref{fig:icecube-cross}.

The DIS curve is calculated from the theoretical formalism of
Ref.\,\cite{nu-dis-cross} by assuming the same mixture of
neutrino and antineutrino events and
by extending our knowledge of the accelerator-based PDFs 
to the higher-energy region.
Within the errors of the IceCube measurements, 
the standard model curve is consistent with the data.
In future, more accurate measurements are expected from 
the IceCube, and there are also future projects of
KM3NeT and Baikal-GVD on ultra-high-energy cosmic
neutrino measurements, so that we should keep our eyes on
these measurements. Since they are in the energy region 
beyond our accelerators, there may be new phenomena 
and new physics beyond the current standard model.

\vspace{-0.10cm}
\subsection{Nuclear modifications}
\label{nuclear-mod}
\vspace{-0.10cm}

As the neutrino measurements become accurate in
the neutrino oscillation experiments, it became necessary
to understand nuclear corrections because neutrino-oxygen 
interactions are involved in the measurements, although
they are partly constrained by near-detector experiments.
For future CP violation measurements for the lepton sector,
neutrino-nucleus cross sections should be calculated
within about 5\% accuracy.
The corrections are generally 20-30\% effects in structure
functions and the PDFs for medium and large nuclei.
They need to be taken into account for a precise cross section estimate.

\begin{figure}[t]
    \hspace{+1.00cm}
 \begin{minipage}[c]{0.45\textwidth}
    \vspace{-0.00cm}
    \hspace{-0.50cm}
    \includegraphics[width=6.3cm]{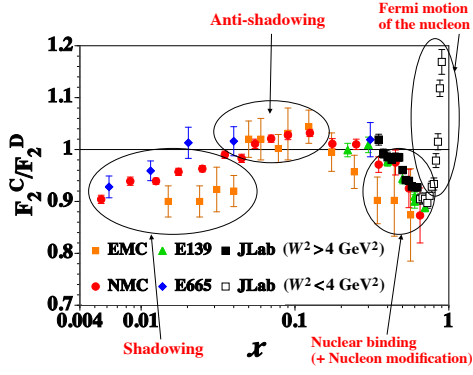}
\vspace{-0.30cm}
\caption{Nuclear modifications in $F_2^C / F_2^D$.}
\label{fig:F2C-D}
\vspace{-0.8cm}
 \end{minipage}
\end{figure}

Now, nuclear modifications of the structure function $F_2$ 
are relatively well known for nuclei with small mass numbers
to large ones in charged-lepton DIS,
although the small $x$ ($<0.004$) region should be yet to 
be investigated by the Electron-Ion Collider (EIC) project.
The typical nuclear modification measurements are shown in 
Fig.\,\ref{fig:F2C-D} for the carbon nucleus as 
the function of the scaling variable $x$.
The open squares indicate the data with small invariant
mass $W^2 < 4$ GeV$^2$, so that they are not DIS data to be precise.
However, the rise of the large-$x$ ratio agrees 
with theoretical expectation based on a convolution model
with nucleon Fermi motion effects, so that they are shown 
together with the DIS data to see the general $x$-dependence
tendency.

At small $x$ ($<0.05$), the modifications are negative and 
they are caused by nuclear shadowing. 
The virtual photon could fluctuate into a $q\bar q$ 
or a vector meson state at small $x$. 
The $q\bar q$ propagation length is estimated as 
$\lambda = 1/|E_V -E_\gamma | = 2 \nu / (Q^2 + M_V^2 ) 
 = 0.2 \ \text{fm} /x >$2 fm at $x<0.1$. 
It becomes larger than the average separation of nucleons in a nucleus,
and multiple scattering occurs. A double-scattering contribution is 
negative to the single scattering one, which results in the shadowing.
Namely, the $q\bar q$ pair or a vector meson strongly interacts with
surface nucleons, so that the projectile does not interact with internal
nucleons and they are shadowed by the surface ones.

The medium- and large-$x$ regions ($x>0.3$) 
of the nuclear modifications are generally described
by the convolution picture. Namely, the nuclear structure function 
$F_2^A$ is calculated by the convolution integral of the nucleonic
structure function $F_2^N$ with the spectral function,
which indicates the nucleon momentum distribution in a nucleus.
The spectral function contains nuclear binding and nucleon 
Fermi-motion effects.
The nuclear binding and Fermi motion are of the order of
10-100 MeV range, which is very small in comparison with
the DIS energies of 10-100 GeV. However, a slight shift
in the nucleon lightcone-momentum distribution results
in the 10\% modifications in the carbon nucleus
through the structure function $F_2^N$.
The Fermi-motion increase with $x$ at large $x$ ($>0.8$)
is also understood in the same theoretical framework
of the convolution description.
Finite nuclear structure functions $F_2^A$ exist in
the kinematical region $0<x<A$ where $A$ is the mass number,
whereas it should vanish for the nucleon at $x=1$.
It means that the ratio becomes $F_2^A / F_2^N \to \infty$
as $x \to 1$, as suggested in Fig.\,\ref{fig:F2C-D}.

The positive modification at $x =0.1$ is called anti-shadowing.
Although such a positive effect should exist according to
the baryon-number, charge, and momentum conservations 
for a nucleus to cancel the negative modifications
of the shadowing and binding, a physics mechanism 
behind the anti-shadowing is not well investigated. 
There is a theoretical suggestion on constructive interference
in the multiple scattering description \cite{brodsky-anti}.
This region is becoming interesting in the sense that
the Miner$\nu$a experiment started producing
nuclear modification data in this region
in neutrino DIS. The Miner$\nu$a data seem to indicate 
large (possibly too large) nuclear modifications
in the region $0.05<x<0.2$ \cite{minerva}.
The Fermilab-E772 Drell-Yan experiment indicated no nuclear
modification in antiquark distributions, so that
the anti-shadowing of $F_2^A$ should be interpreted
by valence-quark modifications.
Currently, the Fermilab-E906 experiment is in progress, so that
new information could be obtained for nuclear antiquark distributions
in this region in the near future.

\begin{figure}[t]
    \hspace{0.550cm}
\begin{minipage}[c]{0.50\textwidth}
    \vspace{0.30cm}
    \hspace{-0.55cm}
     \includegraphics[width=3.90cm]{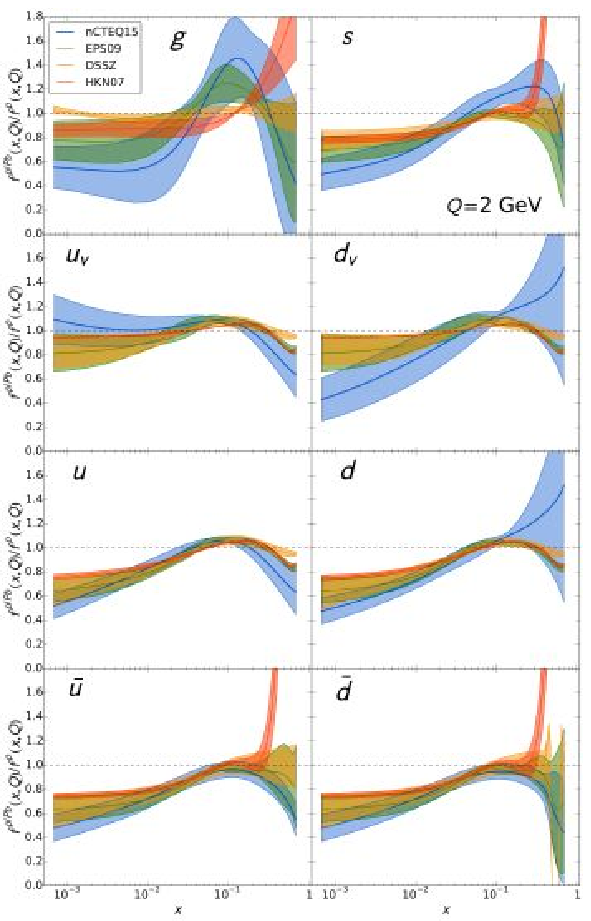}
     \includegraphics[width=3.97cm]{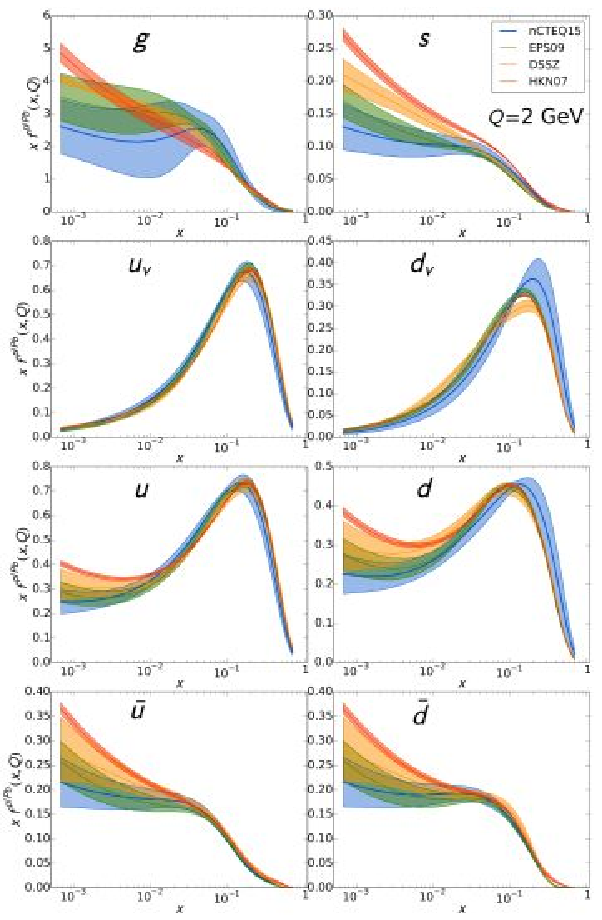} \\
\vspace{-0.60cm}
\caption{Nuclear modifications and NPDFs in iron \cite{ncteq-2016}.}
\label{fig:npdf-pb}
\vspace{-0.80cm}
\end{minipage}
\end{figure}

The nuclear parton distribution functions (NPDFs) are determined
by analyzing experimental data on nuclear structure functions,
Drell-Yan processes, and so on.
Typical results are shown in Fig.\,\ref{fig:npdf-pb} \cite{ncteq-2016},
where the nuclear modifications are shown for each parton distribution
on the left-hand side and
the NPDFs themselves are shown on the right-hand side 
at $Q^2 = 2^2$ GeV$^2$.
Neutrino DIS measurements do not play a major part so far
in ``directly'' finding the nuclear modifications of the PDFs,
although measurements have been done for the heavy targets
such as iron and lead. 
It is because accurate deuteron measurements do not
exist in neutrino reactions, although such data are taken
by the ratio form $F_2^A/F_2^D$ in charged-lepton reactions
as shown in Fig.\,\ref{fig:F2C-D}. If a neutrino factory
with high-intensity neutrino beam will be realized in future,
such data should be obtained also for the deuteron.
In determining the nuclear modifications, such ratio data are desirable. 

Nuclear corrections are often applied 
within neutrino experimental collaborations to publish their
structure functions of ``the nucleon'' in final papers.
Therefore, the neutrino measurements have been useful
in determining the nucleonic PDFs instead of the nuclear PDFs.
They are especially valuable for determining the valence-quark
distribution through the structure function $F_3$ which does
not exist in the charged-lepton DIS.
Neutrino-induced opposite-sign dimuon events are also
important for finding the strange quark distribution
in comparison with the light antiquark distributions:
$s+\bar s \simeq 0.4 (\bar u + \bar d)$
at relatively small $Q^2$.

In Fig.\,\ref{fig:npdf-pb}, nCTEQ analysis results are shown
in comparison with other distributions.
All the analysis results roughly agree within uncertainty bands;
however, the gluon modification is not well determined. 
It could be determined more accurately by using other measurements
such as $J/\psi$ production in the ultra-peripheral
heavy-ion collisions \cite{j/psi-upc}. 
In future, the EIC could provide us accurate information 
on scaling violation of $F_2^A$ at small $x$ $(<0.004)$
to probe the nuclear gluon shadowing accurately.
In addition, there should be progress in future 
on flavor dependent nuclear modifications 
at JLab by the parity-violating DIS and 
at Fermilab by Drell-Yan processes with nuclear targets.

\vspace{-0.00cm}
\section{Gravitational form factors in neutrino scattering
through hadron tomography}
\label{gravitational}
\vspace{-0.00cm}

\begin{wrapfigure}[9]{r}{0.16\textwidth}
    \hspace{-0.55cm}
\begin{minipage}[c]{0.18\textwidth}
    \vspace{-0.30cm}
    \hspace{+0.10cm}
    \includegraphics[width=3.0cm]{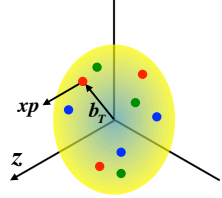}
\vspace{-0.75cm}
\caption{3D structure.}
\label{fig:3d-hadron}
\vspace{-0.5cm}
\end{minipage}
\end{wrapfigure}

We discussed the structure functions and PDFs in Sec.\,\ref{dis-nucleus}
in neutrino DIS. The PDFs indicate distributions of the longitudinal
momentum fraction $x$ of partons. Recently, three dimensional structure
of the nucleon became a hot topic in hadron physics by including
the transverse distributions in addition to the longitudinal PDFs
as illustrated in Fig.\,\ref{fig:3d-hadron}.

\begin{figure}[t]
    \hspace{+0.00cm}
 \begin{minipage}[c]{0.48\textwidth}
    \vspace{-0.00cm}
    \hspace{-0.00cm}
    \includegraphics[width=7.8cm]{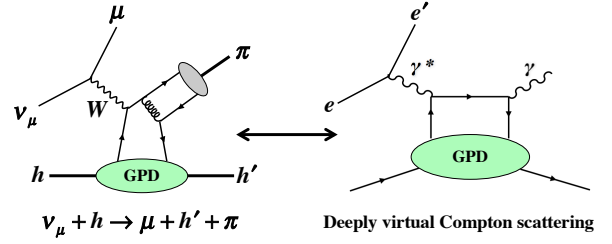}
\vspace{-0.30cm}
\caption{GPDs in $\nu$ reaction and virtual Compton scattering.}
\label{fig:nu-gpds}
\vspace{-0.6cm}
 \end{minipage}
\end{figure}

It was motivated by the topic of solving the origin of nucleon spin.
According to the original quark model of 1964, the nucleon spin
should be interpreted by the combination of three spin-1/2 quark
spins. In other words, the nucleon spin should be carried by
quarks with 100\% probability. However, polarized charged-lepton
DIS measurements indicate that it is a small amount of 20-30\%.
The nucleon spin is one of fundamental physics quantities, so that
it is necessary to find its origin.
The gluon spin may carry a significant fraction of nucleon spin;
however, its contribution has still large uncertainty. 
It will be clarified by the EIC project.
On the other hand, angular momenta of partons could also
contribute to the nucleon spin. Their effects should be
found by one of three dimensional (3D) structure functions,
generalized parton distributions (GPDs), which can be measured
by the virtual Compton scattering process 
on the right-hand side  of Fig.\,\ref{fig:nu-gpds}.
Experimental studies are in progress at 
JLab (Thomas Jefferson National Accelerator Facility)
and CERN-COMPASS. The field of 3D structure functions
is called hadron tomography. Using this technique, 
we can determine gravitational sources, namely gravitational
masses, pressures, and shear forces, in hadrons in terms 
of fundamental quark and gluons, and it is also possible 
in neutrino scattering in principle if its intensity 
is high enough.

The GPDs are defined matrix elements of non-local vector operators,
and their moments are given as
\vspace{-0.10cm}
\begin{align}
 \left ( \frac{P^+}{2}  \right ) ^n
\int dx \, x^{\, n-1} &
\! \! \int\frac{d y^-}{2\pi}e^{i x P^+ y^- /2} \,
  \overline{q}(-y/2) \gamma^+ q(y/2) \Big |_{y^+ = \vec y_\perp =0}
\nonumber \\
&
 = \overline q (0) \gamma^+ 
 \left ( i \overleftrightarrow \partial^+  \right )^{n-1} \!
 q(0) .
\label{eqn:tensor-int-gpd}
\\[-0.70cm]
\nonumber
\end{align}
This operator is the energy-momentum tensor of a quark for $n=2$, 
and it is a source of gravity, whereas
it is the vector-type electromagnetic current for $n=1$
\cite{kst-2018}.
Therefore, measurements of the GPDs can provide information
on the gravitational source within a hadron.
The second moments of the GPDs are expressed by gravitational
form factors. The electromagnetic form factors of the nucleon
have been measured by charged-lepton scattering, and
axial form factors are contained in neutrino-scattering
cross sections. There is a way to access the GPDs, for example,
by the exclusive pion production in neutrino scattering
as shown in Fig.\,\ref{fig:nu-gpds} \cite{nu-gpds}.
The pion production processes have been investigated in
neutrino reactions; however, they are measured so far
at relatively low energies. 
For measuring the GPDs, namely, for the factorization of
the cross section into the soft GPD part and the hard
pQCD one, the kinematical condition,
$Q^2 \gg |t|, \, \Lambda_{QCD}^2$, should be satisfied.
In future, if high-energy and high-intensity neutrino factory
will be realized, such measurements could be possible.

\begin{figure}[b]
 \vspace{-0.20cm}
    \hspace{+0.70cm}
 \begin{minipage}[c]{0.48\textwidth}
    \vspace{-0.00cm}
    \hspace{-0.70cm}
    \includegraphics[width=7.8cm]{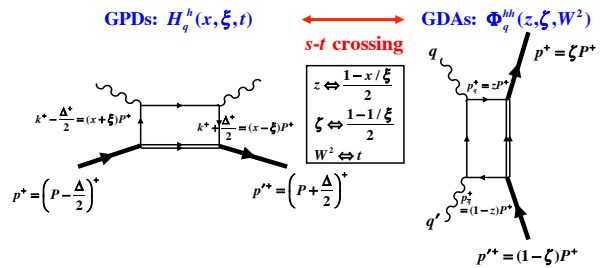}
\vspace{-0.30cm}
\caption{GPDs and GDAs in two-photon process.}
\label{fig:gdas}
\vspace{-0.4cm}
 \end{minipage}
 \vspace{-0.10cm}
\end{figure}

There is another way probe the gravitational form factors by 
timelike processes as shown in Fig.\,\ref{fig:gdas}.
Instead of the virtual Compton scattering for the GPDs, 
the 3D structure functions called 
the generalized distribution amplitudes (GDAs)
can be investigated by the two-photon process
$\gamma^* \gamma \to h \bar h$.
Such measurements were reported by the Belle collaboration
first in 2016 for $\gamma^* \gamma \to \pi^0 \pi^0$ 
\cite{belle-2016}.
The GDAs could be considered as ``timelike GPDs'',
and they are also defined by the same non-local operator
of Eq.\,(\ref{eqn:tensor-int-gpd}). Therefore, the gravitational
form factors can be obtained by analyzing the KEKB measurements.

\begin{figure}[t]
 \vspace{-0.20cm}
\begin{center}
 \begin{minipage}[c]{0.40\textwidth}
    \vspace{-0.00cm}
    \hspace{+0.90cm}
    \includegraphics[width=4.8cm]{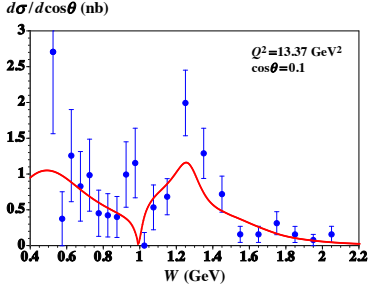}
\vspace{-0.30cm}
\caption{Comparison with Belle cross-section data
of $\gamma^* \gamma \to \pi^0 \pi^0$ for determining the GDAs
\cite{kst-2018}.}
\label{fig:comparison-belle}
\vspace{-0.4cm}
 \end{minipage}
 \end{center}
 \vspace{-0.50cm}
\end{figure}

From the analysis of the two-photon measurements on 
$\gamma^* \gamma \to \pi^0 \pi^0$ of KEKB, 
the GPDs of the pion were obtained \cite{kst-2018}.
There are two gravitational form factors for the spin-0 pion,
and they are calculated by using the determined GDAs.
The GDAs are expressed by a number of parameters, which
are determined by a $\chi^2$ analysis of Belle measurements.
The obtained theoretical cross section is shown in 
Fig.\,\ref{fig:comparison-belle} in comparison with
typical Belle data. Since it is a timelike process,
we should be careful in taking resonances into account
in analyzing the Belle data. In fact, the prominent peak
is clear for $f_2 (1270)$, whereas $f_0 (500)$ and 
$f_0 (980)$ are not so obvious in Fig.\,\ref{fig:comparison-belle}.

\begin{figure}[b]
\begin{center}
 \begin{minipage}[c]{0.40\textwidth}
    \vspace{-0.20cm}
    \hspace{+0.90cm}
    \includegraphics[width=4.8cm]{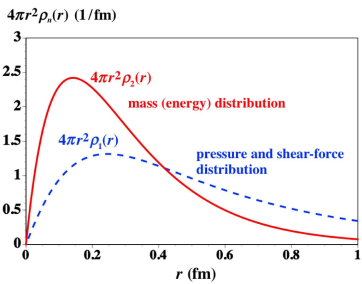}
\vspace{-0.40cm}
\caption{Mass and mechanical distributions obtained
from two gravitational form factors $\Theta_1$ and $\Theta_2$.}
\label{fig:rho-pi}
\vspace{-0.20cm}
 \end{minipage}
 \end{center}
\end{figure}

Once the quark GDAs $\Phi_q^{\pi^0 \pi^0} (z,\,\zeta,\,W^2)$ are 
determined, the form factors are calculated in the following way.
The second moment of a GDA is given by the matrix element
of the quark energy-momentum tensor $T_q^{\mu\nu}$ as
\vspace{-0.10cm}
\begin{align}
\int_0^1  dz \, & (2z -1) \, 
\Phi_q^{\pi^0 \pi^0} (z,\,\zeta,\,W^2) 
\nonumber \\[-0.10cm]
& = \frac{2}{(P^+)^2} \langle \, \pi^0 (p) \, \pi^0 (p') \, | 
     \, T_q^{++} (0) \, | \, 0 \, \rangle ,
\label{eqn:integral-over-z}
\\[-0.60cm]
\nonumber
\end{align}
where $T_q^{\,\mu\nu}$ is defined by
$ T_q^{\,\mu\nu} (x) = \overline q (x) \, \gamma^{\,(\,\mu} 
   i \overleftrightarrow D^{\nu)} \, q (x)$
with the covariant derivative 
$D^\mu = \partial^{\,\mu} -ig \lambda^a A^{a,\mu}/2$.
Here, $g$ is the QCD coupling constant and 
$\lambda^a$ is the SU(3) Gell-Mann matrix.
The right-hand side of Eq.\,(\ref{eqn:integral-over-z})
can be expressed by the timelike gravitational form factors 
$\Theta_1$ and $\Theta_2$ of the pion as
\vspace{-0.00cm}
\begin{align}
& \langle \, \pi^0 (p) \, \pi^0 (p') \, | 
     \, T_q^{\mu\nu} (0) \, | \, 0 \, \rangle 
\nonumber \\
& \ \ \ \ \ 
  = \frac{1}{2} 
  \left [ \, \left ( s \, g^{\,\mu\nu} -P^{\,\mu} P^\nu \right ) \, 
  \Theta_{1, q} (s) + \Delta^\mu \Delta^\nu \, 
   \Theta_{2, q} (s) \, \right ] ,
\label{eqn:emt-ffs-timelike-0}
\\[-0.60cm]
\nonumber
\end{align}
where $P$ and $\Delta$ are given by the pion momenta as
$P=p+p'$ and $\Delta=p'-p$.
In this way, the timelike gravitational form factors
are obtained form the Belle measurements.
Next, the timelike form factors are converted to 
the spacelike ones by using the dispersion relation,
and then they are Fourier transformed to become
the spacial mass and mechanical distributions shown
in Fig.\,\ref{fig:rho-pi}.

From the spacial distributions in Fig.\,\ref{fig:rho-pi},
the root-mean-square radii of the pion can be calculated as 
$\sqrt {\langle r^2 \rangle _{\text{mass}}} = 0.69 \, \text{fm}$ and
$\sqrt {\langle r^2 \rangle _{\text{mech}}} = 1.45 \, \text{fm}$,
which correspond to gravitational-mass and mechanical radii.
There are some ambiguities in our analysis in assigning phase factors,
so that the radius estimates have some uncertainty
ranges as \cite{kst-2018}:
$ \sqrt {\langle r^2 \rangle _{\text{mass}}} 
      =  0.56 \sim 0.69 \, \text{fm}$, 
$\sqrt {\langle r^2 \rangle _{\text{mech}}} 
     = 1.45 \sim 1.56 \, \text{fm}$.
This is the first result on the gravitational radii 
from the actual analysis of experimental measurements.
The charge radius of the pion has been already measured as
$\sqrt {\langle r^2 \rangle _{\text{charge}}} =0.672 \pm 0.008$ fm
\cite{pdg-2018}.

\vspace{-0.10cm}
\section{Summary}
\label{summary}
\vspace{-0.10cm}

High-energy neutrino-nucleon and nucleus scattering processes are
interesting and practically important for future precision
neutrino oscillation measurements and for finding a possible 
new phenomenon in ultra-high-energy cosmic neutrino measurements.
For application to the neutrino oscillation experiments, we need
to combine the neutrino DIS cross section with the ones of
other kinematical regions, namely quasi-elastic, resonance, 
and Regge regions, because neutrino measurements are done
at the beam energies form several hundred MeV to several GeV. 
On the ultra-high-energy side, the IceCube collaboration started
publishing neutrino cross sections in the region of $10^{13}$-$10^{15}$ eV,
so that new phenomenon may be observed in future at the energies
beyond the current artificial lepton accelerators.
If a high-intensity neutrino factory will be realized in future,
neutrino scattering measurements could probe gravitational 
form factors of hadrons by exclusive processes through
the studies of 3D tomography. It is currently a hot topic
in the hadron-physics community by using virtual Compton scatting
and two-photon processes for measuring the GPDs and GDAs.

\vspace{-0.20cm}
\section*{Acknowledgments}
\vspace{-0.10cm}

Figures 2 and 5 are used with the copyright permission 
of American Physical Society and authors.
The author thanks M. Bustamante for supplying Figure 3.



\vspace{-5.00cm}

\end{document}